\documentclass{article}
\usepackage{graphicx} % Required for inserting images
\usepackage{amsmath}
\usepackage{xspace}
\usepackage{amssymb}
\usepackage{times}
\usepackage{color}
\usepackage{pifont}
\usepackage{subcaption}
\usepackage{enumitem}
\usepackage{isotope}
\usepackage{upgreek}
\usepackage[T1]{fontenc}
\usepackage[utf8]{inputenc}
\usepackage{wrapfig}
\usepackage{pdfpages}
\usepackage{tocloft}
\usepackage[margin=1in]{geometry}

\usepackage{doi}

\usepackage{url}

\usepackage{hyperref}
\hypersetup{
 colorlinks=true,     % color the words instead of use a colored box
 urlcolor=blue,       % \href{...}{...} external (URL)
 linkcolor=blue,     % \ref{...} and \pageref{...}
 citecolor=blue,     % \cite{}
 plainpages=false,
 breaklinks=true,
 bookmarksnumbered=true,
 bookmarksopen=true,
 pdfpagemode=UseOutlines  % None, UseThumbs, UseOutlines, FullScreen
 }
\usepackage{lineno}
\title{Overview of Neutrino-Mass Determination}
\author{Diana S. Parno
for the KATRIN collaboration}
\date{%
    Department of Physics, Carnegie Mellon University, Pittsburgh PA 15213, USA
}

\begin{document}

\maketitle

\textit{Presented at the 32nd International Symposium on Lepton Photon Interactions at High Energies, Madison, Wisconsin, USA, August 25-29, 2025.}

\section{Introduction}
\label{sec:intro}
Neutrino mass is one of the most enduring open questions of the Standard Model of Particle Physics. Neutrino-oscillation experiments~\cite{Super-Kamiokande:1998kpq,SNO:2002tuh, ParticleDataGroup:2024cfk} have established that the three active neutrino-flavor states are quantum superpositions of three distinct neutrino-mass eigenstates, but these experiments have no sensitivity to the absolute neutrino-mass scale -- only the splittings between values. Major new experiments (DUNE~\cite{abi:2020deep}, Hyper-Kamiokande~\cite{Hyper-Kamiokande:2018ofw}, JUNO~\cite{JUNO:2015zny,JUNO:2021vlw}) aim to establish whether the mass ordering of neutrinos has a similar pattern to other particle families. However, neutrinos are at least a million times lighter than the next lightest particle (the electron), and the mechanism by which they acquire mass is yet unknown.

The question of the neutrino-mass scale is important not only for particle theory, but also for cosmology: as the most abundant known matter particles in the early universe, neutrino mass has shaped the formation of cosmological structures. Cosmic-microwave-background data, in combination with cluster counts and baryon acoustic oscillations, support tight limits on the sum of neutrino-mass values, $\Sigma m_i$. For example, DESI has combined baryon-acoustic-oscillation data with Planck cosmic-microwave-background data to set a limit of $\Sigma m_i < 0.064$~eV (95\%)~\cite{DESI:2025zgx}, when the data are interpreted in the context of the $\Lambda$ Cold Dark Matter ($\Lambda$CDM) model. Thus far, cosmological data have not supported a measurement of the neutrino mass, and indeed seem to prefer an unphysical, negative value for the sum of the masses -- one of a few current tensions in the standard cosmological model. (See, e.g., Refs.~\cite{Abdalla:2022yfr} and~\cite{CosmoVerseNetwork:2025alb} for recent reviews and discussions.) 

A laboratory measurement of the neutrino-mass scale, independent of cosmological assumptions or specific mechanisms for acquiring mass, is therefore a crucial test both for particle theory and for cosmology. In Sec.~\ref{sec:kin-meth}, I will briefly summarize current experimental efforts toward such a test, before going into detail about the KATRIN experiment (Sec.~\ref{sec:katrin-exp}) and its recent results (Sec.~\ref{sec:katrin-results}).

\section{Kinematic Methods for Measuring the Neutrino-Mass Scale}
\label{sec:kin-meth}

Direct, kinematic methods for measuring the neutrino-mass scale, recently reviewed by Formaggio, de Gouv{\^e}a, and Robertson~\cite{Formaggio:2021nfz}, use nuclear $\upbeta$ decay as a laboratory. The extreme high-energy tail of the $\upbeta$ spectrum is shifted and distorted by the presence of non-zero neutrino mass, which represents energy that the $\upbeta$ cannot carry away. In the quasi-degenerate regime, in which the absolute mass scale is large compared to the splittings between states, these measurements are sensitive to the incoherent sum of the neutrino-mass eigenvalues $m_i$, each weighted by the PMNS matrix element $U_{ei}$ that gives its contribution to the electron flavor created in the decay. Often denoted $m_\beta$, this observable is given by
\begin{equation}
    m_\beta = \sqrt{\sum_{i=1}^{3}|U_{ei}|^2m_i^2}
\end{equation}
The great strength of kinematic neutrino-mass measurements is their model independence: the only fundamental assumption made in this analysis is that energy is conserved in $\upbeta$ decay. However, these experiments are extremely challenging technically. Only a small fraction\footnote{About one part in $10^{12}$ for tritium experiments.} of the spectrum carries useful information about the $m_\beta$, so experiments must precisely measure the energy and rate near the spectral endpoint $E_0$ while avoiding being swamped by a vast number of low-energy $\upbeta$ decays.

To improve the statistical sensitivity of a neutrino-mass search, an isotope with a low $Q$-value is preferred: the less kinetic energy is released in the decay, the larger the fraction of the $\upbeta$ spectrum that is sensitive to $m_\beta$. Tritium (${}^3$H or T) has $Q=18.6$~keV and a half-life (12.3~yr) that facilitates handling and storage, and further benefits from the fact that the decay is super-allowed. Molecular tritium, which decays according to
\begin{equation}
    \mathrm{T}_2 \rightarrow {}^3\mathrm{He} + e^- + \bar{\nu}_e,
\end{equation}
has been used by the KATRIN experiment, which will be discussed in more detail in Sec.~\ref{sec:katrin-exp} and~\ref{sec:katrin-results}, to set  
the current world-leading limits on $m_\beta$. Building on neutrino-mass experiments from the 1990s and early 2000s, KATRIN transports $\upbeta$s from tritium decay outside of the source, and performs an integral energy measurement.

Cyclotron resonance emission spectroscopy (CRES)~\cite{Monreal:2009za} is an appealing alternative measurement technique for gaseous tritium sources, with active research and development efforts from the Project~8~\cite{Project8:2017nal} and QTNM~\cite{Amad:2024jod} collaborations. The essential idea is to trap $\upbeta$ electrons in a magnetic field of strength $B$. The frequency $f_c$ of the emitted cyclotron radiation is then
\begin{equation}
    f_c = \frac{1}{2\pi} \frac{eB}{m_e + E_{kin}/c^2}
\end{equation}
Here, $e$ and $m_e$ are the electron charge and mass; $E_{kin}$ is its kinetic energy, which is needed to reconstruct the spectrum; and $c$ is the speed of light. With this measurement technique, the $\upbeta$ need not be transported outside the source where it is created, but the radiated power is strikingly low -- about 1~fW at the endpoint of the tritium spectrum, in a 1~T magnetic field. Despite these challenges, Project~8 has achieved a first CRES-based search for the neutrino-mass scale, based on 3770 T$_2$ decay events measured over 82~live days in a small (1 mm$^3$) waveguide. The resulting limit of $m_\beta < 180$~eV (90\%~C.L.) is a milestone for the CRES technique~\cite{Project8:2022hun, Project8:2023jkj}. Both Project~8 and QTNM are now pushing towards CRES measurements in larger volumes, and towards the realization of atomic tritium sources that would avoid an energy-resolution limit imposed by molecular broadening.

The HOLMES and ECHo experiments~\cite{Alpert:2025tqq, ECHo:2025ook} probe the neutrino-mass scale using the electron-capture decays of ${}^{163}$Ho:
\begin{equation}
    {}^{163}\mathrm{Ho} \rightarrow {}^{163}\mathrm{Dy}^* + \nu_e
\end{equation}
The spectrum of de-excitation energy from this decay is more complex in shape than for $\upbeta$-emitting decays like T$_2$, and requires different tools to measure. With no single $\upbeta$ carrying away a substantial fraction of the kinetic energy, an experiment must capture many small possible contributions to the spectrum, from de-excitation photons to shake-off electrons. However, the endpoint spectral region carries information about $m_\beta$ in the same way that the tritium spectrum does~\cite{DeRujula:1982qt}, and the very low ${}^{163}$Ho Q-value of 2.83~keV~\cite{ECHo:2015qgh} is statistically favorable. 

To measure the de-excitation spectra following ${}^{163}$Ho decay, both ECHo and HOLMES produce ${}^{163}$Ho in a nuclear reactor; embed it in gold microcalorimeters; and measure temperature changes to determine the deposited energy. Using transition-edge sensors to probe temperature, HOLMES has measured some 70M decays, resulting in an upper limit of $m_\beta < 27$~eV (90\% C.L.)~\cite{Alpert:2025tqq}. ECHo, meanwhile, has set an upper limit of $m_\beta < 15$~eV (90\% C.L.) based on 200M decays measured with metallic magnetic calorimeter technology for sensing temperature~\cite{ECHo:2025ook}. These results demonstrate important strides toward both scalability and precision in a challenging nuclear system that tests the effective mass of electron neutrinos, rather than antineutrinos.

\section{The KATRIN Experiment}
\label{sec:katrin-exp}

The KArlsruhe TRItium Neutrino (KATRIN) experiment~\cite{Lokhov:2022zfn}, first conceived in 2001~\cite{KATRIN:2001ttj}, has been taking tritium data since 2019 and has set the world's current best limits on the kinematic neutrino mass $m_\beta$. The KATRIN measurement principle, fully described in Ref.~\cite{KATRIN:2021dfa}, is based on a windowless, gaseous tritium source with $\upbeta$ energy analysis by the principle of Magnetic Adiabatic Collimation with Electrostatic filtering (MAC-E filtering)~\cite{Beamson:MacEfilter1980, Picard:1992kra}. $\upbeta$s are produced via T$_2$ decay in a strong magnetic field, and are then adiabatically guided along magnetic field lines to a region of weak magnetic field (the ``analyzing plane'' of the experiment's main spectrometer). In this adiabatic motion, the magnetic moment is conserved:
\begin{equation}
    \mu = \frac{E_\perp}{B} = \mathrm{const},
\end{equation}
where $B$ is the local magnetic-field strength and $E_\perp$ is the kinetic energy due to motion perpendicular to that field. The overall effect is to rotate the $\upbeta$ momenta, so that the $\upbeta$ flux at the analyzing plane is a broad electron beam, roughly collimated in the longitudinal direction. The application of a longitudinal retarding potential $U$ thus sets a threshold on the total kinetic energy of the $\upbeta$: $E_\upbeta > qU$, where $q$ is the electron charge.  In effect, the KATRIN main spectrometer acts as an integrating high-pass filter, where electrons above $qU$ are transmitted to the detector to be counted, and electrons below the threshold are reflected back. The filter width is set by the ratio of the minimum to the maximum magnetic fields, dictating the degree to which the electrons are collimated; further details are given by Kleesiek et al.~\cite{Kleesiek:2018mel}. KATRIN scans the integrated spectrum by counting transmitted $\upbeta$s at up to 40 threshold set points $qU_i$, spanning an energy range of $E_0 - 300~\mathrm{eV} \leq qU_i \leq E_0 + 135\mathrm{eV}$ over approximately three hours per scan.

\begin{figure}
    \centering
    \includegraphics[width=0.8\linewidth]{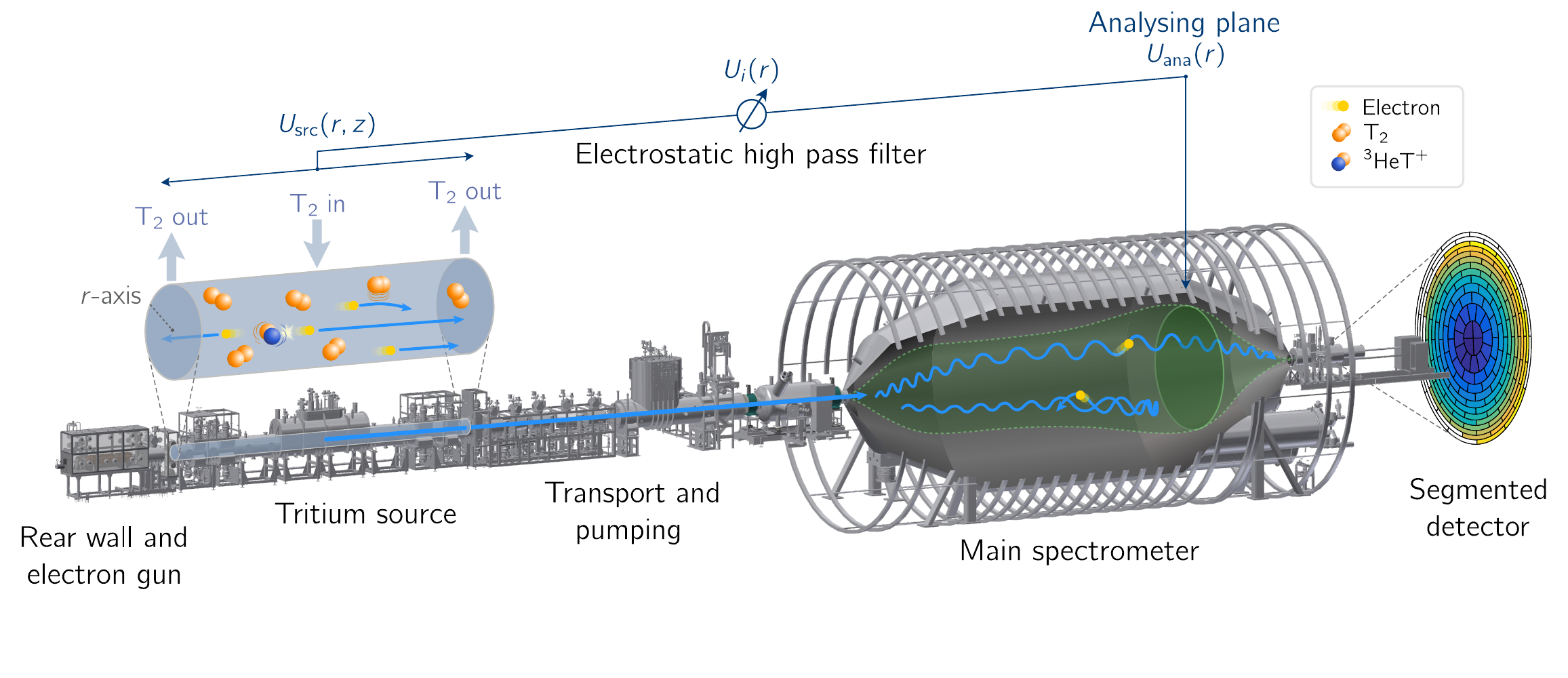}
    \caption{Engineering drawing of the 70-m KATRIN beamline, illustrating the dynamics of the gaseous tritium source (left). The visualization of the main spectrometer includes a drawing of the location of the shifted analyzing plane, as well as a transmitted (top) and reflected (bottom) $\upbeta$. Reproduced with permission from Ref.~\cite{KATRIN:2024cdt}.}
    \label{fig:beamline}
\end{figure}

As shown in Fig.~\ref{fig:beamline}, the main spectrometer is part of an extended KATRIN beamline 70~m in length. The windowless, gaseous tritium source provides some $10^{11}$ decays per second, with an atomic purity of more than 95\% purity as verified by laser Raman spectroscopy~\cite{KATRIN:LARA}. Upstream of the source, a calibration section contains X-ray monitors and an electron gun for measuring the column density, the amount of gas that a $\upbeta$ must traverse en route to the detector. Downstream of the source, differential and cryogenic pumping stages remove tritium gas while preserving adiabatic transport of the electrons. A small MAC-E filter -- the pre-spectrometer -- was initially used to pre-filter $\upbeta$s by energy, but is now left at a minimal retarding potential to mitigate backgrounds. The 1240~m$^3$ main spectrometer is followed by the detector section, which guides $\upbeta$s to a monolithic, segmented silicon p-i-n diode.

To extract $m_\beta^2$ from the data, the measured spectrum is fit to a model spectrum $R_{calc}$, which depends on the theoretical tritium $\upbeta$ spectrum $R_\beta$ and the experimental response function $f_{calc}$. At each scan step $qU_i$, the calculated spectral rate is given by

\begin{equation}
    \label{eq:calc-rate}
    R_{calc}(qU_i) = A_s N_{\mathrm{T}}\int_{qU_i}^{\infty} R_\beta\left(E; E_0, m_\beta^2\right) f_{calc}(E, qU_i)dE + R_{bg}.
\end{equation}
The four free parameters of this model fit are a normalization factor $A_s$; a flat background rate $R_{bg}$; the spectral endpoint $E_0$; and the square of the effective neutrino mass, $m_\beta^2$. The theoretical spectrum $R_\beta\left(E; E_0, m_\beta^2\right)$ incorporates molecular final-state effects and the known physics of tritium $\upbeta$ decay. The response function $f_{calc}(E, qU_i)$, determined through calculation and through detailed systematics measurements with an electron gun and a gaseous ${}^{83m}$Kr calibration source among other configuration changes, models the probability that a $\upbeta$ of energy $E$ will be transmitted through the apparatus and counted at the detector; energy loss due to scattering in the source is a major contribution to this function. $N_T$ gives the number of tritium atoms in the source, weighted by the detection efficiency and the geometric acceptance of the MAC-E filter. Systematic uncertainties are treated as nuisance parameters in the fit. Details of the analysis scheme may be found in Refs.~\cite{KATRIN:2024cdt} and~\cite{KATRIN:2021fgc}.

Due to the MAC-E filter technique, KATRIN's primary backgrounds are quite unusual. By design, signal $\upbeta$s are decelerated to almost zero kinetic energy at the analyzing plane of the main spectrometer; once they have crossed this region, they are then symmetrically re-accelerated to ground. Low-energy electrons created just downstream of the analyzing plane are therefore energetically indistinguishable from the signal. KATRIN's electromagnetic fields effectively shield against secondary electrons emitted from the spectrometer walls (induced, e.g., by environmental gamma radiation~\cite{KATRIN:2019dnj} or by cosmic-ray muons~\cite{KATRIN:2018rxw}), so the primary remaining background is from electrons created within the main-spectrometer volume.

During KATRIN's construction, the main spectrometer was exposed to basement air for several years. Although a clean-room environment was maintained, ${}^{222}$Rn circulated in this air. After several steps in the decay chain, ${}^{214}$Po sometimes $\alpha$-decayed on the vessel wall, giving enough of a kick to the decay product ${}^{210}$Pb to implant it in the wall. After two additional $\upbeta$ decays, the decay chain reaches another $\alpha$ emitter: ${}^{210}$Po. The recoil from this decay releases highly excited (large-$n$) atoms from the stainless-steel surface, including ${}^{206}$Pb; H; Fe; and O. These neutral Rydberg atoms easily penetrate the electromagnetic shielding to diffuse through the volume of the main spectrometer, where some are ionized by blackbody radiation and others are autoionized. This rather convoluted background mechanism has been confirmed as KATRIN's dominant background via several tests, including a temporary contamination with ${}^{220}$Rn; this decay chain proceeds quite similarly to ${}^{222}$Rn, but all the decay products are short-lived~\cite{KATRIN:2020zld}. 

It is impractical to remove this ``Rydberg background'', but it can be mitigated. The background is produced when a Rydberg atom is ionized downstream of the analyzing plane; prior to ionization, these atoms are distributed randomly throughout the volume. KATRIN can therefore reduce the background by shrinking the imaged volume. By using magnetic air-core coils already present as compensators for external magnetic fields, the analyzing plane can be shifted much closer to the detector~\cite{Lokhov:2022iag}. In this operational mode with a ``shifted analyzing plane'' (SAP; contrasted with the normal analyzing plane, or NAP), the energy resolution is worsened, and the analysis is more complex. Notably, due to greater field variation across the analyzing surface~\cite{KATRIN:2024qct}, different sections or ``patches'' of the segmented detector must be analyzed using slightly different transmission functions. However, this mode reduces the background by a factor of 2.

\section{KATRIN Results}
\label{sec:katrin-results}

KATRIN's measurement campaigns are denoted ``KNM$x$'', where KNM stands for KATRIN Neutrino Mass and $x$ is a sequential number. Table~\ref{tab:katrin-ops} summarizes basic conditions for the five measurement campaigns that are part of KATRIN's latest published neutrino-mass results. 

\begin{table}[]
    \centering
    \begin{tabular}{l|llll}
        \textbf{Campaign} & \textbf{Time Period} & \textbf{Analyzing Plane} & \textbf{MTD} & \textbf{Notes} \\
        \hline
        \hline
        KNM1 & Spring 2019 & Normal & Nominal & Low source activity, $2.5\times 10^{10}$~Bq \\
        & & & & Result: $m_\beta^2 = \left(-1.0^{+0.9}_{-1.1}\right)$~eV$^2$ \\
        & & & & \qquad\qquad $m_\beta < 1.1$~eV (90\% C.L.)~\cite{KATRIN:2019yun} \\         
        \hline
        KNM2 & Fall 2019 & Normal & Nominal & Nominal source activity, $9.5\times 10^{10}$~Bq\\
        & & & & Results: $m_\beta^2 = \left(0.26\pm0.34\right)$~eV$^2$ \\
        & & & & \qquad \qquad $m_\beta < 0.9$~eV (90\% C.L.)~\cite{KATRIN:2021uub}.\\
        \hline
        KNM3-SAP & Summer 2020 & Shifted & Nominal &  \\
        KNM3-NAP & Summer 2020 & Normal & Nominal &  \\
        \hline
        KNM4-NOM & Fall 2020 & Shifted & Nominal & Elevated backgrounds \\
        KNM4-OPT & Fall 2020 & Shifted & Optimized & \\
        \hline
        KNM5 & Spring 2021 & Shifted & Optimized \\
        \hline
    \end{tabular}
    \caption{Summary of KATRIN campaigns used in published neutrino-mass analyses. The \textit{Analyzing Plane} column refers to either the central, ``normal'' position, or the downstream ``shifted'' position to mitigate backgrounds (Sec.~\ref{sec:katrin-exp}). The \textit{MTD} column refers to the Measurement Time Distribution, or the amount of time spent at each scan; some campaigns used the original, ``nominal'' MTD, while others used a modified, ``optimized'' MTD. All neutrino-mass limits are quoted at a 90\% confidence level (C.L.). Some numbered campaigns are subdivided due to changes in operational conditions.}
    \label{tab:katrin-ops}
\end{table}

A neutrino-mass analysis involves fitting a model spectrum (Eq.~\ref{eq:calc-rate}) to the data, with $A_s, R_{bg}, E_0,$ and $m_\beta^2$ as free parameters, and with systematic uncertainties captured by nuisance parameters. As described in Ref.~\cite{KATRIN:2021fgc}, the computational burden of the fit is reduced by several simplifications.
Data from detector pixels with similar transmission functions are grouped together; for data acquired with the normal analyzing plane, all detector pixels can be grouped together, but analysis of shifted-analyzing-plane data requires subdivision into 13 ``patches''. During a campaign, data from like scan steps $qU_i$ are stacked together so that many scans can be analyzed as one spectrum. The collaboration maintains two separate modeler/fitter packages for frequent cross-checks. 

A two-level blindness scheme mitigates experimenter's bias. The analysis methods are first fixed using simulated data sets, generated using the measured operational parameters of the real runs. Second, after cross-checks, the first fits to the data are model-blinded. The molecular final states excited by $\upbeta$ decay in T$_2$ distort the $\upbeta$ spectrum in a similar way to $m_\beta^2$. The theoretical $\upbeta$ spectrum in Eq.~\ref{eq:calc-rate}, $R_\beta(E; E_0, m_\beta^2)$ uses as input an ab initio calculation of this molecular final-state distribution (FSD)~\cite{Saenz:2000dul,Schneidewind:2023xmj}. If the model spectrum is instead generated with an alternate FSD, its variance smeared by some unknown amount, then the fit result is blind to the actual $m_\beta^2$ signature. 

The analysis accounts for the variation of experimental conditions (Table~\ref{tab:katrin-ops}) via a joint fit  across campaigns. The operating parameters are allowed to vary separately across campaigns, but $m_\beta^2$ is held as a common fit parameter. Meanwhile, a model spectrum is predicted for each campaign and for each detector patch. The overall fit for the KNM1--5 data incorporates 1609~data points across all spectra, with 144~correlated systematic parameters.

For the data from KNM1--5, the uncertainty budget is dominated by a $1\sigma$ statistical uncertainty of 0.108~eV$^2$, with the total systematic contribution coming to 0.072~eV$^2$~\cite{KATRIN:2024cdt}. Compared to previous KATRIN results, uncertaintes related to backgrounds are substantially reduced, driven by the implementation of the shifted analyzing plane and by the decision to mostly de-energize the pre-spectrometer. The collaboration is continuing to actively improve its understanding of other systematics. 

After the KNM1--5 data were first unblinded, the collaboration discovered two issues in the analysis. First, the KNM4 campaign contained runs that were too different to combine into one spectrum: the change in measurement-time distribution, combined with a high-voltage drift over time, caused an unexpected bias when the data were stacked together. This was addressed by splitting KNM4 into the KNM4-NOM and KNM4-OPT campaigns, distinguished by the measurement-time distribution. Second, an inconsistency was found between the results from two different methods of measuring the $\upbeta$ energy loss due to scattering in the source. The uncertainty of the energy-loss model was re-estimated to account for this discrepancy, and the collaboration is now working to finalize a new model that reconciles the two measurement methods. After these issues were discovered, the collaboration paused the analysis and reviewed every systematic uncertainty in teams before launching the unblinding process again.

\begin{wrapfigure}{L}{0.6\textwidth}
    \centering
    \includegraphics[width=0.8\linewidth]{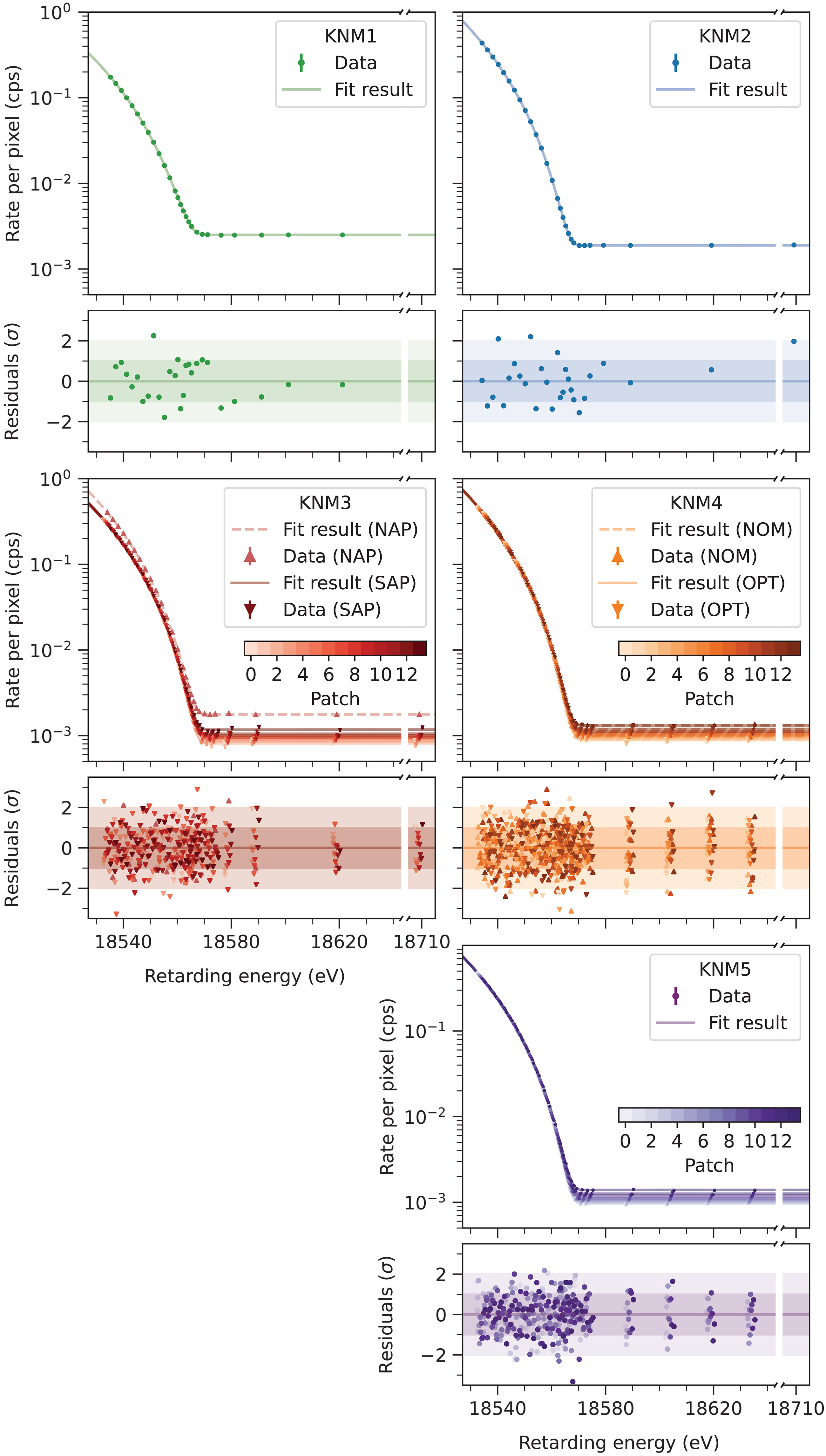}
    \caption{Results of a joint maximum-likelihood fit to all KNM1--5 data. For each campaign, the top panel displays the rate data (markers) and fit result (lines) at each scan step $qU_i$, while the bottom panel displays the residuals in units of $\sigma$. For measurements with a shifted analyzing plane, data are separated out by patch, or region of the detector. Reproduced with permission from Ref.~\cite{KATRIN:2024cdt}.}
    \label{fig:spectra}
\end{wrapfigure}

Figure~\ref{fig:spectra} shows the results of a joint maximum-likelihood fit to all KNM1--5 data, with $p = 0.84$. Campaigns in the normal-analyzing-plane (NAP) operational mode display single spectra, with data combined across the entire detector; shifted-analyzing-plane (SAP) campaigns show 13 spectra, one for each patch of the detector. 

The best-fit $m_\beta^2$ value,
\begin{equation}
    m_\beta^2 = 0.14^{+0.13}_{-0.15}~\mathrm{eV}^2,
\end{equation}
is slightly negative, plausible within statistical fluctuations. Negative best-fit results are expected for any experiment measuring a value that turns out to be below its sensitivity: when the truth value is zero, a fit to the data will fluctuate negative approximately half the time. KATRIN constructs an upper limit on the neutrino-mass scale using the Lokhov-Tkachov method~\cite{Lokhov:2014zna}. This method is conceptually quite similar to the Feldman-Cousins method~\cite{Feldman:1997qc}, but for a negative fluctuation in the best-fit value, it uses the sensitivity as an upper limit rather than allowing a negative fluctuation to tighten the upper limit. The resulting upper limit is
\begin{equation}
    m_\beta < 0.45~\mathrm{eV}~(90\%~ \mathrm{C.L.}),
\end{equation}
a better than fourfold improvement on the sensitivities of the previous generation of experiments~\cite{Kraus:2004zw, Troitsk:2011cvm}. These results are published in detail in Ref.~\cite{KATRIN:2024cdt}.

\section{Discussion and Outlook}
\label{sec:discussion}

KATRIN's latest results derive from data acquired from Spring~2019 through Spring~2021. The experiment has continued to acquire neutrino-mass data since that time, with its most recent scan occurring in October~2025; the collaboration is targeting a sensitivity of $m_\beta < 0.3~\mathrm{eV}~(90\%~ \mathrm{C.L.})$. At present, the collaboration is refining its measurements of systematic uncertainties while preparing to re-configure the beamline for a deep spectral measurement, far beyond the endpoint region. With lower $\upbeta$ energy comes a dramatic increase in rate, necessitating a novel, 1500-pixel silicon drift detector -- called TRISTAN~\cite{KATRIN:2018oow, Urban:2021ink}. Additional planned beamline changes also include modifications to the source region, which will run at a lower activity to control pileup, and to the electromagnetic field configuration to preserve adiabatic transport across a wider range of energies.

A precise measurement of the tritium $\upbeta$ spectrum allows searches for a wide range of beyond-Standard-Model (BSM) phenomena. By extending the spectral measurement past the endpoint region of neutrino mass, KATRIN can probe different characteristic mass scales for such BSM signatures. For example, hypothesized sterile neutrinos -- which do not interact via the weak force, but which do mix with the known active-flavor neutrinos via a fourth mass eigenvalue $m_4$ -- are in principle visible in a $\upbeta$ spectrum as a distinctive kink-shaped structure, corresponding to an increase in phase space below $E_0 - m_4$ where this mass state can contribute to the electron flavor state created in the decay. Using the KNM1--5 data, KATRIN~\cite{KATRIN:2025lph} has ruled out a large portion of the parameter space favored for a sterile-neutrino solution to the gallium anomaly~\cite{Barinov:2021asz}, and entirely excludes the Neutrino-4 claim~\cite{Serebrov:2023wqf} of a sterile-neutrino discovery in an oscillation experiment. Using early data sets with low tritium activity, KATRIN set competitive limits on $m_4$ in the keV range, which is appealing as a warm dark matter candidate~\cite{KATRIN:2022spi}. After the Phase~II upgrade with the TRISTAN detector, KATRIN will extend its sensitivity to lower mixing angles and higher mass values.

Beyond sterile neutrinos, KATRIN's first two measurement campaigns have been used to set limits on Lorentz-invariance violation in the neutrino sector~\cite{KATRIN:2022qou}, on generalized neutrino interactions~\cite{KATRIN:2024odq}, and on a local overdensity of relic neutrinos~\cite{KATRIN:2022kkv}. Additional avenues for BSM searches are described in Ref.~\cite{KATRIN:2022ayy}, and the full KATRIN data set will allow much improved precision over any early results.

In this phase and the next, KATRIN has achieved world-leading precision on tritium $\upbeta$ spectroscopy, providing kinematic tests of cosmology, particle theory, and oscillation-based sterile neutrino searches. As KATRIN fully grapples with both increased statistics and hardware upgrades, many exciting results are yet to come. 

\section*{Acknowledgments}
    We acknowledge the support of the Helmholtz Association (HGF), Ministry for Education and Research BMBF (05A23PMA, 05A23PX2, 05A23VK2, and 05A23WO6), the doctoral school KSETA at KIT, Helmholtz Initiative and Networking Fund (grant agreement W2/W3-118), Max Planck Research Group (MaxPlanck@TUM), and Deutsche Forschungsgemeinschaft (DFG) (GRK 2149 and SFB-1258 and under Germany’s Excellence Strategy EXC 2094–390783311) in Germany; the Ministry of Education, Youth, and Sport (CANAM-LM2015056, LTT19005) in the Czech Republic; Istituto Nazionale di Fisica Nucleare (INFN) in Italy; the National Science, Research, and Innovation Fund through the Program Management Unit for Human Resources \& Institutional Development, Research and Innovation (grant B37G660014) in Thailand; and the US Department of Energy through awards DE-FG02-97ER41020, DE-FG02-94ER40818, DE-SC0004036, DE-FG02-97ER41033, DE-FG02-97ER41041, DE-SC0011091, and DE-SC0019304 and the Federal Prime Agreement DE-AC02-05CH11231 in the United States. This project has received funding from the European Research Council (ERC) under the European Union Horizon 2020 research and innovation program (grant agreement 852845). We thank the computing cluster support at the Institute for Astroparticle Physics at Karlsruhe Institute of Technology, Max Planck Computing and Data Facility (MPCDF), and the National Energy Research Scientific Computing Center (NERSC) at Lawrence Berkeley National Laboratory.

\end{document}